\documentclass[a4paper]{jpconf}
\usepackage{graphicx}
\usepackage{bm}
\usepackage{amsmath}
\usepackage{color}

\newcommand{\kv}{{\bm k}}

\newcommand{\nv}{{\bm n}}

\newcommand{\qv}{{\bm q}}
\newcommand{\rv}{{\bm r}}

\newcommand{\beqa}{\begin{eqnarray}}
\newcommand{\eeqa}{\end{eqnarray}}

\newcommand{\vare}{\varepsilon}

\begin{document}
\title{Spin Hall Current and Spin-transfer Torque in Ferromagnetic Metal}

\author{Junya Shibata}

\address{Kanagawa Institute of Technology, 
1030 Shimo-Ogino Atsugi, Kanagawa 243-0292, Japan}

\ead{shibata@gen.kanagawa-it.ac.jp}

\author{Hiroshi Kohno}
\address{
Graduate School of Engineering Science, Osaka University,
Toyonaka, Osaka 560-8531, Japan}%
\ead{kohno@mp.es.osaka-u.ac.jp}

\begin{abstract}
We theoretically examine the spin-transfer torque 
in the presence of spin-orbit interaction (SOI) 
at impurities in a ferromagnetic metal 
on the basis of linear response theory. 
We obtained, in addition to the usual spin-transfer torque, 
a new contributioin 
$\sim {\bm j}_{\rm SH}^{\phantom{\dagger}} \!\cdot\! \nabla {\bm n}$
in the first order in SOI, 
where ${\bm j}_{\rm SH}^{\phantom{\dagger}}$ is the spin Hall current 
driven by an external electric field.  
This is a reaction to inverse spin Hall effect 
driven by spin motive force in a ferromagnet. 
\end{abstract}

\section{Introduction}

Since the theoretical proposal by Slonczewski \cite{Slonczewski96} 
and Berger \cite{Berger96} of the current-induced magnetization reversal 
in nanopillars 
and subsequent experimental realizations, 
the spin current has been recognized to be useful in 
manipulating magnetization in tiny magnets. 
An important concept is the spin-transfer torque 
that the spin current exerts on magnetization. 
 For a smooth spin texture ${\bm n}$, it is expressed as 
$(\hbar/2e)\, {\bm j}_{\rm s} \!\cdot\! \nabla {\bm n}$ \cite{BJZ98,KTS06}. 
 Here $e>0$ is the elementary charge, 
${\bm j}_{\rm s}=\sigma_{\rm s}{\bm E}$ is the spin-current 
density driven by an applied electric field ${\bm E}$, and 
$\sigma_{\rm s} = \sigma_{\uparrow}-\sigma_{\downarrow}$ 
is the ``spin conductivity'' with 
$\sigma_{\uparrow}$ $(\sigma_{\downarrow})$ 
being the diagonal conductivity for majority-spin (minority-spin) 
electrons.

Recently, we theoretically studied the spin Hall current 
\cite{Hirsch99} in a ferromagnetic conductor 
in the presence of spin texture ${\bm n}$ and SOI at impurities 
\cite{SK09}. 
 It was shown that an electric field ${\bm E}$ also induces a 
spin Hall current, 
${\bm j}_{\rm SH}^{\phantom{\dagger}} 
 = \sigma_{\rm SH}^{\phantom{\dagger}} {\bm n} \times {\bm E}$, 
where 
$\sigma_{\rm SH}^{\phantom{\dagger}} 
 = \sigma_{{\rm H}\uparrow}^{\phantom{\dagger}} 
 + \sigma_{{\rm H}\downarrow}^{\phantom{\dagger}}$ 
is the spin Hall conductivity 
with $\sigma_{{\rm H}\uparrow}^{\phantom{\dagger}}$ 
$(\sigma_{{\rm H}\downarrow}^{\phantom{\dagger}})$ 
being Hall conductivity for majority-spin (minority-spin) electrons 
\cite{note0}. 
 The total spin current ${\cal J}_{\rm S}$ is thus given by 
\beqa
\label{js}
{\cal J}_{\rm S} 
= {\bm j}_{\rm s} + {\bm j}_{\rm SH}^{\phantom{\dagger}} 
= \sigma_{\rm s}{\bm E} + \sigma_{\rm SH}^{\phantom{\dagger}} 
 {\bm n} \times {\bm E} . 
\eeqa
 Naturally, this spin Hall current (${\bm j}_{\rm SH}^{\phantom{\dagger}}$) 
is expected to contribute to the spin-transfer torque as 
\beqa
\label{sht}
{\bm t}_{\rm el}^{\rm H} 
= 
\frac{\hbar}{2e} \, 
({\bm j}_{\rm s} + {\bm j}_{\rm SH}^{\phantom{\dagger}}) 
\cdot\nabla\nv . 
\eeqa

 The spin Hall conductivity $\sigma_{\rm SH}^{\phantom{\dagger}}$ 
in Eq.~(\ref{js}) is given by the correlation function between 
spin current and charge current, and its reciprocal effect 
(in the sense of Onsager) described by the same function is the 
inverse spin Hall effect which is driven by spin motive force 
(SMF) \cite{{Berger86},{Stern92},{BM07},{Saslow07},{Duine08}, 
{Tserkovnyak08},{YBKXNTE09}} or 
a spin-dependent effective electric field 
${\bm E}_{\rm s}$ \cite{{Duine08},{Tserkovnyak08}}, 
and induces a charge current 
\beqa
\label{j}
{\cal J}
= \sigma_{\rm s}{\bm E}_{\rm s} 
 + \sigma_{\rm SH}^{\phantom{\dagger}} {\bm n} \times {\bm E}_{\rm s} .
\eeqa
 Such a spin-dependent field ${\bm E}_{\rm s}$ is known to arise from the 
dynamics of textured magnetization, and is given by 
$E_{{\rm s},i} = (\hbar/2e) \, {\bm n} \cdot 
 (\partial_{i}{\bm n}\times \dot{{\bm n}})$. 
 The first term of Eq.(\ref{j}) is actually 
the reciprocal effect of the ordinary spin-transfer effect 
(first term of Eq.~(\ref{sht})) \cite{Duine08}.

 The purpose of this paper is to derive the second term of Eq.~(\ref{sht}) 
microscopically and clarify the relation 
to the corresponding SMF (second term of Eq.~(\ref{j})), 
both of which arise in the presence of SOI.

\section{Model and Calculation}

 We consider the $s$-$d$ model with conducting $s$ electrons 
and localized $d$ spins, ${\bm n}$, which are coupled via the 
$s$-$d$ exchange interaction, $-M {\bm n} \cdot {\bm \sigma}$. 
 The $s$ electrons are subjected to impurity potential 
$V_{\rm imp}({\bm r})= u\sum_{i}\delta({\bm r}-{\bm R}_{i})$, 
where $u$ and ${\bm R}_{i}$ are the strength 
and position of the impurity, as well as SOI, 
$\sim i\lambda_{\rm so} {\bm \sigma} \cdot (\nabla \times V_{\rm imp})$, 
at impurities \cite{SK09}. 
In order to treat electrons in a spin texture, 
we perform a local transformation 
in electron spin space and take the spin quantization axis 
to be the $d$-spin direction $\nv$ at 
each point of space and time \cite{{KMP77},{Volovik87},{TF94}}. 
 The Lagrangian in the rotated frame is given by \cite{note1}
\beqa
\label{L_el}
&&L_{\rm el} = \int d\rv~ a^{\dagger}(x)
\left[i\hbar
\frac{\partial}{\partial t}+
\frac{\hbar^2}{2m}\nabla^{2}
+\vare_{\rm F}
+M \sigma^{z}-V_{\rm imp}(\rv)
\right]a(x),
\label{Lagrangian-e}\\
&&\tilde{H}_{\rm so}= \lambda_{\rm so} \frac{m}{\hbar}
\vare_{ij\alpha}\int d\rv~
(\partial_{i}V_{\rm imp}(\rv))
{\cal R}^{\alpha\beta}(x)\tilde{j}^{\beta}_{j}(x),\\
&&H_{\rm e-A} = 
\int d\rv~\tilde{j}^{\alpha}_{i}(x)A^{\alpha}_{i}(x) .
\eeqa
 Here 
$a^{\dagger}(x)=(a^{\dagger}_{\uparrow}(x),a^{\dagger}_{\downarrow}(x))$ 
is the electron creation operator at $x=({\bm r},t)$ in the rotated frame, 
$\sigma^{\alpha}$'s are Pauli matrices, 
$\vare_{\rm F}$ is the Fermi energy, 
$\lambda_{\rm so}$ is the strength of SOI, 
$\vare_{ij\alpha}$ is the complete antisymmetric tensor 
with $\vare_{xyz}=1$, 
${\cal R}^{\alpha\beta} = 2m^{\alpha}m^{\beta}-\delta^{\alpha\beta}$ 
is a $3\times 3$ orthogonal matrix (with 
${\bm m}=(\sin(\theta/2) \cos\phi, \sin(\theta/2)\sin\phi, 
\cos(\theta/2))$ 
for 
${\bm n}=(\sin\theta \cos\phi, \sin\theta \sin\phi, 
\cos\theta)$), 
$\tilde{j}^{\alpha}_{i} = (\hbar/2mi)
(a^{\dagger}\sigma^{\alpha}\partial_{i}a
-\partial_{i}a^{\dagger}\sigma^{\alpha}a)$ 
is the spin-current density operator 
and 
$A^{\alpha}_{i}= ({\bm m}\times\partial_{i}{\bm m})^{\alpha}$ 
is the SU(2) gauge field. 
 Repeated indices imply summation over $i,j,\alpha=x,y,z$.

In the $s$-$d$ model, the spin torque is generally given by \cite{KS07}
\beqa
{\bm t}_{\rm el}(x) = M 
 {\bm n}(x) \times {\cal R}(x)
\langle \tilde{{\bm \sigma}}(x) \rangle_{\rm ne}, 
\label{torque}
\eeqa
where 
$\langle \tilde{{\bm \sigma}}(x)\rangle_{\rm ne}=
\langle a^{\dagger}(x) {\bm \sigma} a(x) \rangle_{\rm ne}$ 
is the spin polarization of $s$ electrons evaluated 
in an appropriate non-equilibrium state. 
For current-induced torques, it is calculated as a linear response 
to ${\bm E}$ as 
\beqa
\langle \tilde{\sigma}^{\mu}(\qv)\rangle_{\rm ne} = 
\lim_{\omega \to 0}
\frac{
\chi^{\mu}_{i}(\qv,\omega+i0)
-
\chi^{\mu}_{i}(\qv,0)
}{
i\omega
}E_{i}, 
\eeqa
where $\chi^{\mu}_{i}(\qv,\omega)$ is the current-spin 
correlation function obtained, for example, through an analytic 
continuation from the Matsubara representation 
\beqa
\chi^{\mu}_{i}(\qv,i\omega_{\lambda})
=
\int_{0}^{1/T}d\tau 
e^{i\omega_{\lambda}\tau}
\langle {\rm T}_{\tau}\tilde{\sigma}^{\mu}(\qv,\tau)J_{i}\rangle .
\eeqa
Here $T$ is the temperature, 
which will be eventually set to zero, 
$\omega_{\lambda} = 2\pi \lambda T$ with $\lambda$ being an integer, 
and $J_i$ is the current operator.

 In this paper, we focus on the skew-scattering process 
(in the terminology of anomalous Hall effect \cite{{DCB01}})
and neglect the side-jump process. 
 This corresponds to taking as the current operator 
$J_i \simeq \tilde{J}_{i} 
 \equiv -e\sum_{\kv} (\hbar k_{i}/m) a^{\dagger}_{\kv}a_{\kv}$
neglecting the anomalous velocity term due to SOI. 
 In the lowest order in $\lambda_{\rm so}$, 
the first contribution to $\chi^{\mu}_{i}$ comes from 
the third-order impurity scattering 
\cite{note2} and is first order in both 
$\tilde{H}_{\rm so}$ and $H_{\rm e-A}$. 
 The diagrammatic expressions are shown in Fig. 1. 
\begin{figure}
\begin{center}
\includegraphics[width=36pc]{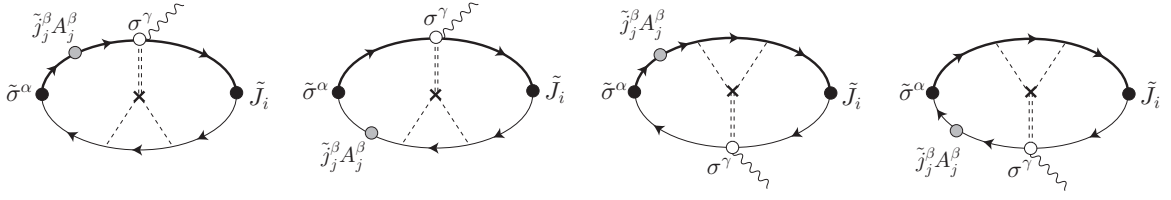}
\end{center}
\caption{Feynman diagrams for the coefficients, $\chi^{\mu}_{i}$, 
of the spin torque induced by spin Hall current due to skew scattering. 
 The thick (thin) solid line represents an electron line 
carrying Matsubara frequency $i\vare_{n}+i\omega_{\lambda}$ $(i\vare_{n})$. 
 The dotted line (double dotted line with an open circle) represents potential 
(spin-orbit) scattering $V_{\rm imp}$ ($\tilde{H}_{\rm so}$) by impurities. 
 The gray circle represents the interaction with 
the SU(2) gauge field. 
 The wavy line represents the rotation matrix ${\cal R}^{\alpha\beta}$.}
\end{figure}
After some calculations, we obtain 
\beqa
\chi^{\mu}_{i}(\qv,i\omega_{\lambda}) 
&=& -ien_{\rm i}u^{3}\lambda_{\rm so}\frac{4}{9\hbar}
\vare_{ij\lambda}
\sum_{\qv'}n^{\lambda}_{\qv-\qv'}A^{\nu}_{j}(\qv')\nonumber\\
&&\times
T\sum_{n}
\sum_{\sigma}
\sigma(\delta^{\mu\nu}_{\perp}+i\sigma\vare^{\mu\nu}_{\perp})
\left\{
J_{3}^{\sigma}
J_{2}^{\sigma}
I_{1}^{\sigma}
-\left(
J_{3}^{\bar{\sigma}}
J_{2}^{\bar{\sigma}}
I_{1}^{\bar{\sigma}}
\right)^{*}
\right\}, 
\eeqa
where 
$n^{\alpha}_{\qv}$ is the Fourier component of the spin texture
$n^{\alpha}(\rv)$
(note that $n^{\alpha}={\cal R}^{\alpha z}$), 
$\sigma = \uparrow, \downarrow$ corresponds, respectively, 
to $\sigma = +1, -1$ in the formula 
(and to $\bar{\sigma}=\downarrow, \uparrow$ or $-1, +1$), 
$\delta^{\mu\nu}_{\perp}=\delta^{\mu\nu}-\delta^{\mu z}\delta^{\nu z}$, 
$\vare^{\mu\nu}_{\perp}=-\vare^{\nu\mu}_{\perp}$ with $\vare^{xy}_{\perp}=1$. 
We have defined 
$J_{3}^{\sigma}
=\sum_{\kv}G^{+}_{\kv\sigma}G^{+}_{\kv\sigma}G_{\kv\sigma}$, 
$J_{2}^{\sigma}
=\sum_{\kv}G^{+}_{\kv\sigma}G_{\kv\sigma}$, 
$I_{1}^{\sigma}
=\sum_{\kv}G^{+}_{\kv\sigma}$, 
where 
$G_{\kv\sigma}(z) = (z-\vare_{\kv} + \vare_{{\rm F}\sigma}
+i\gamma_{\sigma}{\rm sgn}({\rm Im}z))^{-1}$ 
and 
$G^{+}_{\kv\sigma} \equiv G_{\kv\sigma}(i\vare_{n}+i\omega_{\lambda})$ 
are the impurity-averaged thermal Green's functions, 
with $\vare_{\kv}=\hbar^{2}\kv^2/2m$, 
$\vare_{{\rm F}\sigma}= \vare_{F}+\sigma M$, 
$\gamma_{\sigma} = \hbar/2\tau_{\sigma}
=\pi n_{\rm i}u^{2}\nu_{\sigma}$ 
and
$\nu_{\sigma}=(m/2\pi^{2}\hbar^{2})
\sqrt{2m\vare_{{\rm F}\sigma}}/\hbar$.

After analytic continuation, 
$i\omega_{\lambda} \to \omega + i0$, we 
evaluate as $J^{\sigma}_{3}=(\sigma/2M)
3\pi n^{\rm el}_{\sigma}\tau_{\sigma}$, 
$J^{\sigma}_{2}= 3\pi n^{\rm el}_{\sigma}$ and 
$I^{\sigma}_{1}= -i\pi\nu_{\sigma}$
($n^{\rm el}_{\sigma}=(2/3)\nu_{\sigma}\vare_{{\rm F}\sigma}$) 
in the lowest order of 
$\gamma_{\sigma}/\vare_{{\rm F}\sigma}$ 
and 
$\gamma_{\sigma}/M$. 
 Then we obtain   
\beqa
\chi^{\mu}_{i}(\qv,\omega+i0)-\chi^{\mu}_{i}(\qv,0)
=i\omega\frac{1}{M} \frac{\hbar}{e} \, 
 \sigma_{\rm SH}^{\phantom{\dagger}} \, 
 \vare_{ij\lambda}
\sum_{\qv'}n^{\lambda}_{\qv-\qv'} \, \delta^{\mu\nu}_{\perp}
A^{\nu}_{j}(\qv'),
\eeqa
where 
$\sigma_{\rm SH}^{\phantom{\dagger}} 
=\sigma^{\rm skew}_{\uparrow}
+\sigma^{\rm skew}_{\downarrow}$ 
is the spin Hall conductivity with 
\beqa
\sigma^{\rm skew}_{\uparrow(\downarrow)}= 
\lambda_{\rm so}u\frac{2\pi e^{2}}{\hbar}
(n^{\rm el}_{\uparrow(\downarrow)})^{2} \, 
\tau_{\uparrow(\downarrow)}
\eeqa
Thus the spin polarization is given by 
\beqa
\langle \tilde{\sigma}^{\mu}(x)\rangle_{\rm ne}
=\frac{1}{M}\frac{\hbar}{e} \, j_{{\rm SH},i}^{\phantom{\dagger}}(x) \, 
\delta^{\mu\nu}_{\perp}A^{\nu}_{j}(x), 
\label{spin}
\eeqa
where 
${\bm j}_{\rm SH}^{\phantom{\dagger}}(x) 
 = \sigma_{\rm SH}^{\phantom{\dagger}} \nv(x)\times {\bm E}$ 
is the spin Hall current density \cite{SK09}. 
Substituting Eq. (\ref{spin}) into Eq. (\ref{torque}) and 
 using the relation, 
${\cal R}^{\gamma\mu}\delta^{\mu\nu}_{\perp}A^{\nu}_{i}
=-\left(\nv\times\partial_{i}\nv\right)^{\gamma}/2$ \cite{KS07}, 
we obtain 
\beqa
t^{{\rm H},\alpha}_{\rm el}(x) &=& M 
\vare_{\alpha\beta\gamma}n^{\beta}(x){\cal R}^{\gamma\mu}(x)
\frac{1}{M}\frac{\hbar}{e} \, {\bm j}_{\rm SH}^{\phantom{\dagger}}(x) \, 
 \delta^{\mu\nu}_{\perp}A^{\nu}_{i}(x)\nonumber\\
&=& \frac{\hbar}{2e} \, {\bm j}_{\rm SH}^{\phantom{\dagger}}(x) 
 \!\cdot\! \nabla n^{\alpha}(x). 
\label{torque2}
\eeqa
This is the desired spin-transfer torque due to spin Hall current. 
 Combining with the ordinary spin-transfer torque, 
$(\hbar/2e) \, {\bm j}_{\rm s} \!\cdot\! \nabla \nv$ \cite{KTS06}, 
and Eq. (\ref{torque2}), 
the total spin-transfer torque in a disordered ferromagnetic metal 
is given by Eq.~(\ref{sht}) with Eq.~(\ref{js}) .

\section{Summary and discussion}

 We have shown that the spin-transfer torque 
in the presence of SOI at impurities has a contribution from the 
spin Hall current, given by the second term of Eq.(\ref{sht}). 
 The result is consistent with the picture that it is a reciprocal 
effect to the corresponding SMF (second term of Eq.~(\ref{j})), 
since both are characterized by the same 
coefficient $\sigma_{\rm SH}^{\phantom{\dagger}}$.

 We have considered only the skew-scattering process, but not 
the side-jump process. 
 Calculation of the latter process seems quite complicated and 
less straightforward compared to the former, in that the anomalous current 
is absent at the left (spin) vertex in the diagram, as opposed to the case of 
anomalous Hall conductivity. 
 The whole calculation including the side-jump process will be reported 
in a future publication \cite{SK09_2}.

\ack
This work is supported by a Grant-in-Aid from Monka-sho, Japan.


\begin{thebibliography}{9}
\bibitem{note0}
Note that each Hall conductivity contributes 
similarly to the spin Hall conductivity 
since SOI curves the orbits of majority-spin and 
minority-spin electrons in mutually oppsite directions.
\bibitem{note1}
We have suppressed terms which are unrelated 
to the present calculation. 
For the full Lagrangian, please see Ref. \cite{SK09}.
\bibitem{note2}
In the present calculation, 
we take the average for the random potential as 
$\overline{V_{\rm imp}(\rv_{1})V_{\rm imp}(\rv_{2})V_{\rm imp}(\rv_{3})}
=n_{\rm i}u^{3}\delta(\rv_{1}-\rv_{3})\delta(\rv_{2}-\rv_{3})$, 
where $n_{\rm i}$ is the impurity concentration.
\end{thebibliography}
\end{document}